\documentclass[prd,11pt]{revtex4-1}

\usepackage{amsmath}
\usepackage{amsfonts}
\usepackage{amssymb}
\usepackage{color}
\usepackage{graphicx}
\usepackage[all]{xy}
\usepackage{color}
\usepackage[T1]{fontenc}
\usepackage{lmodern}
\usepackage[utf8]{inputenc}
\usepackage{graphicx}
\usepackage[titletoc,title]{appendix}
\begin{document}

\title{Classical and quantum cosmology with two perfect fluids: stiff matter and radiation}
\author{F. G. Alvarenga}
\email{flavio.alvarenga@ufes.br}
\affiliation{Departamento de Ci\^encias Naturais, CEUNES, Universidade Federal do Esp\' \i rito Santo, CEP 29933-415, S\~ao Mateus, ES,
Brazil.}
\author{R. Fracalossi}
\email{rfracalossi@gmail.com}
\affiliation{Departamento de F\' \i sica, CCE, Universidade Federal do Esp\' \i rito Santo, CEP 29075-910, Vit\'oria, ES, Brazil.}
\author{R. C. Freitas}
\email{rodolfo.camargo@pq.cnpq.br}
\affiliation{Departamento de F\' \i sica, CCE, Universidade Federal do Esp\' \i rito Santo, CEP 29075-910, Vit\'oria, ES, Brazil.}
\author{S. V. B. Gon\c{c}alves}
\email{sergio.vitorino@pq.cnpq.br}
\affiliation{Departamento de F\' \i sica, CCE, Universidade Federal do Esp\' \i rito Santo, CEP 29075-910, Vit\'oria, ES, Brazil.}

\date{\today}

\begin{abstract}
In this work the homogeneous and isotropic Universe of Friedmann-Robertson-Walker is studied in the presence of two fluids: stiff matter and radiation described by the Schutz's formalism. We obtain to the classic case the behaviour of the scale factor of the universe. For the quantum case the wave packets are constructed and the wave function of the universe is found.
\end{abstract}

\maketitle

PACS number(s): 98.80.-k, 98.80.Cq, 98.80.Qc \vspace{0.7cm}

\section{Introduction}

Quantum cosmology \cite{Israel, Wiltshire} is one of the aspects explored in the quantum gravity program. Its application to the Universe as a whole takes place at very early times the universe was small indeed. For exemple, in the Planck time, about $10^{-44}$ s after the Big Bang,  its size was the order of $10^{-33}$ cm suggesting that quantum effects dominated \cite{Atkatz}. In general the quantum cosmological models are built considering a finite number of degrees of freedom in the presence of geometrical symmetry inherent to the universe. The quantization of a system subjected to these conditions is called quantization in a minisuperspace. Furthermore, with the quantum cosmology it is possible to study a real specific system in which quantum effects of gravity are fundamental. As the early Universe is the best of laboratories where different quantum theories of gravitation can be tested, quantum cosmology has as one of its motivations to serve as a powerful auxiliary tool in the construction of a final and fundamental quantum theory of gravity.
\par
Considering only the cosmological point of view, quantum cosmology plays an important role in solving a serious problem of the standard cosmological model: The existence of an initial singularity. Mechanisms such as quantum tunneling used in more simple cosmological models or in sophisticated approaches such as volume quantization provided by the loop formalism are examples of consistent solutions of this problem \cite{Bojowald}. In addition, the theory allows us to establish the initial conditions for inflation, for primordial perturbations and for spontaneous symmetry breaking \cite{Konstantin,Pad}. That is, the quantum cosmology is a theory of initial conditions.
\par
Minisuperspace quantization is a reasonable approximation to describes certain quantum gravitational effects. It is expected however that the inclusion of other degrees of freedom in the problem must refine the results obtained and lead to a more accurate description of the phenomenon studied, even lead to additional technical complications. This addition of new degrees of freedom can be done in several different ways. In the hydrodynamic description of the material content, consistent and obvious way to refine the description of the Universe is to use not just one but two fluids \cite {Fracalossi, Tovar}.
\par
In this paper we choose stiff matter ($\alpha = 1$) and radiation ($\alpha = 1/3$) as matter content, which play an important role in the early Universe. The energy density of this fluid in the cosmological gauge ($N(t) = 1$) is proportional to $1/a(t)^4$. As the energy density of the stiff matter \cite{Zeldovich}, in the same gauge, is proportional to $1/a(t)^6$, there must have been a time before the radiation in which the stiff matter dominated. In fact, the abundance of particles produced after the Big Bang due to expansion and cooling of the Universe is an implication of the presence of this fluid with $\alpha = 1$ at this stage \cite{Kamionkowski, Joyce, Salati, Pallis, Gomez, Pallis2, Germano}.
\par
The material content is treated here according to the Schutz formalism \cite{Schutz1, Schutz2}. With this description we can make the degrees of freedom of the fluid plays the role of time in the theory, transforming the Wheeler-DeWitt equation, which governs the dynamic behaviour of the system, in a Schr\"{o}dinger-like equation whose solution is a time-dependent wave function, among other variables. So the complex problem of the absence of a variable linked to the time evolution in quantum gravity is solved giving to time a purely phenomenological character leading to a well-defined Hilbert space structure.

This paper is organized as follows: Section (\ref{class}) presents the Schutz's formalism in a classical FRW cosmological model with stiff matter and radiation. The analytical evolution of the scale factor of the universe, $a(t)$, is obtained. We emphasize that, since this point, time is introduced following the Schutz's formalism. We present next, Section (\ref{quant}), the quantum model and we obtain the equation that governs the dynamics of the model, the Wheeler-DeWitt equation, which is solved. A detailed discussion and summary of these results is provided in Section (\ref{concl}).

\section{Classical FRW model with stiff matter and radiation described by the Schutz's formalism}
\label{class}

We consider a homogeneous and isotropic Universe whose geometry is described by the FRW metric, which is given by
\begin{equation}
ds^2=-N(t)^2dt^2+a(t)\left(\frac{1}{1-kr^2}+r^2d\theta^2+r^2sin^2\theta d\phi^2\right)\;\;\;,\label{frw}
\end{equation}
where $N(t)$ is the lapse function, $k$ is the space section curvature, which can assume the values $-1,0,1$ and $a(t)$ is the scale factor of the Universe.

The action associated to the gravitational field is given by
\begin{equation}
S_g=\int \sqrt{-g}Rd^4x+\int_{\partial M}d^3x\sqrt{h}h_{ij}K^{ij}\;,\label{acao}
\end{equation}
in units such that $(8\pi G/3)^{\frac{1}{2}}=1$, $R$ is the scalar of curvature of spacetime, $h_{ij}$ is the 3-metric on the boundary  $\partial M$ of the 4-manifold $M$ and $K_{ij}$ is the extrinsic curvature. Replacing the metric (\ref{frw}) in (\ref{acao}) and using the ADM formalism of general relativity \cite{ADM}, it's possible to write the gravitation Hamiltonian like
\begin{equation}
H_g=-\frac{p_a^2}{24a}-6ka\;.\label{hg}
\end{equation}

We consider in this model a Universe whose material content is described by two fluids: stiff matter and radiation. To describe the dynamics of these fluids, we use the Schutz's formalism  \cite{Schutz1,Schutz2}. The matter content will consist of a perfect fluid with pressure $p$, density of total mass-energy $\rho$, and specific enthalpy $\mu$. In this formalism, the 4-velocity is described from five velocity-potentials
\begin{equation}
U_{\nu}=\frac{1}{\mu}(\phi_{,\nu}+\theta S_{,\nu}+\varphi\beta_{,\nu})\;,\label{4vel}
\end{equation}
where only the specific entropy $S$ has any direct physical interpretation. Moreover, $\varphi$ and $\beta$ are associated to rotational movements and are nulls in homogeneous and isotropic models. This 4-velocity is normalized like
\begin{equation}
g_{\alpha\beta}U^{\alpha}U^{\beta}=-1\;,\label{unorm}
\end{equation}
and with this is possible to write the enthalpy in terms of the potentials
\begin{equation}
\mu=\frac{1}{N}(\dot{\phi}+\theta\dot{S})\,\,\label{entalpia2} .
\end{equation}

The Schutz's representation is associated to a variational principle whose action is especially simple
\begin{equation}
S=\int_M d^4x\sqrt{-g}p\;.\label{smat}
\end{equation}
The pressure of the perfect fluid can be obtained from the barotropic equation of state $p=\alpha\rho$. The basic thermodynamic relations of the fluid take the form \cite{Rubakov}
\begin{equation}
\rho=\rho_0(1+\Pi)\, ;\, \quad \mu=(1+\Pi) + \frac{p}{\rho_{0}} \, ; \, \quad T_fdS=d\Pi +p d\bigg(\frac{1}{\rho_{0}}\bigg) \, ,  \,
\end{equation}
where $T_{f}$ is the temperature and $\Pi$ is the specific internal energy.

If we write 
\begin{equation}
d\Pi +p d\bigg(\frac{1}{\rho_{0}}\bigg)=(1+\Pi)d[ln(1+\Pi)-\alpha\,ln\rho_0]\,\, ,
\end{equation}
we can identify $T_f=1+\Pi\,\, \mbox{and} \,\, S=ln(1+\Pi)/{\rho_{0}}^{\alpha}$. 

After a few mathematical steps, the energy density and pressure reduces to
\begin{equation}
\rho=\left(\frac{\mu}{\alpha+1}\right)^{1+\frac{1}{\alpha}}e^{-\frac{S}{\alpha}}\,\,\, \mbox{and}\,\,\, p=\alpha\left(\frac{\mu}{\alpha+1}\right)^{1+\frac{1}{\alpha}}e^{-\frac{S}{\alpha}}\,\, \label{pressure}.
\end{equation}

If we use  (\ref{entalpia2}) and (\ref{pressure})  in the matter action (\ref{smat}), it is possible to obtain the Lagrangian of the fluid
\begin{equation}
L_{f}=N^{-\frac{1}{\alpha}}a^3\,\frac{\alpha}{(\alpha+1)^{1+\frac{1}{\alpha}}}(\dot{\phi}+\theta\dot{S})^{1+\frac{1}{\alpha}}\, e^{-\frac{S}{\alpha}}\, .\label{fk9}
\end{equation}

The fluid conjugated momenta are derived from the above Lagrangian, written in terms of the canonical variables
\begin{equation}
p_{\phi}=N^{-\frac{1}{\alpha}}a^3\, (\alpha+1)^{-\frac{1}{\alpha}}\,(\dot{\phi}+\theta\dot{S})^{\frac{1}{\alpha}}\, e^{-\frac{S}{\alpha}} \, ,
\end{equation}
\begin{equation}
p_{S}=\frac{\partial L_{f}}{\partial \dot{S}} = \theta\, p_{\phi}\, ,
\end{equation}
and
\begin{equation}
p_{\theta}=0\, .
\end{equation}

By means of the Legendre transformation
\begin{equation}
L_{f}=\dot{\phi}p_{\phi} + \dot{S}p_{S} - N{\cal H}_{f}\,\, ,
\end{equation}
we obtain the matter Hamiltonian
\begin{equation}
\label{hf1}
H_f=\frac{p_{\phi}^{\alpha+1}e^S}{a^{3\alpha}}\;.
\end{equation}

This can be put in a more suggestive form by means of the canonical transformations
\begin{equation}
T=-p_S e^{-S}p_{\phi}^{-(\alpha+1)}\; ,\label{tr1}
\end{equation}
\begin{equation}
{\overline{\phi}}=\phi-(\alpha+1)\frac{p_S}{p_{\phi}}\; , \label{tr2}
\end{equation}
\begin{equation}
p_T=p_{\phi}^{(\alpha+1)}e^S\;,\label{ptr1}
\end{equation}
and
\begin{equation}
{\overline{p}}_{\phi }=p_{\phi}\;.\label{ptr2}
\end{equation}

With this, the equation (\ref{hf1}) is written in the form
\begin{equation}
H_f=\frac{p_T}{a^{3\alpha}}\;.\label{hf2}
\end{equation}

So, the total Hamiltonian will be
\begin{displaymath}
\mathcal{H}=\mathcal{H}_g+\mathcal{H}_f\; ,
\end{displaymath}
\begin{equation}
\label{htotal}
\mathcal{H}=-\frac{p_a^2}{24a}-6ka+\frac{p_T}{a^{3\alpha}}\;.
\end{equation}
Note that the Hamiltonian is linear in one of the momenta and it's possible to introduce the variable $T$ as global phase time \cite{Hajicek}.

The ADM formalism also shows that the lapse function $N$ acts like a Lagrange multiplier of the system so that
\begin{equation}
\mathcal{H}\approx 0\;,\label{vinc}
\end{equation}
where the symbol $\approx$ which means zero weakly. 

We now consider two non-interacting perfect fluids. Then, we can write
\begin{equation}
H=N\mathcal{H}=N\left(-\frac{p_a^2}{24a}-6ka+\frac{p_T}{a^{3\alpha}}+\frac{p_{\sigma}}{a^{3\beta}}\right)\;.\label{Hf}
\end{equation}
At this point, we particularize to the radiation fluid, ($\alpha=\frac{1}{3}$) and stiff matter ($\beta=1$) and we consider the case of null curvature ($k=0)$ so that the equation (\ref{Hf}) can be write in the form
\begin{equation}
H=N\mathcal{H}=N\left(-\frac{p_a^2}{24a}+\frac{p_T}{a}+\frac{p_{\sigma}}{a^{3}}\right)\;.\label{Hf001}
\end{equation}
Using the Hamilton equations, we can write
\begin{equation}
\dot{p_a}=-\frac{\partial(N\mathcal{H})}{\partial a}=-N\left(\frac{p_a^2}{24a^2}-\frac{p_T}{a^2}-3\frac{p_{\sigma}}{a^4}\right)\;,\label{h1}
\end{equation}
\begin{equation}
\dot{p_T}=-\frac{\partial(N\mathcal{H})}{\partial T}=0\;\;\rightarrow\;\; p_T=const.\;,\label{h2}
\end{equation}
\begin{equation}
\dot{p_{\sigma}}=-\frac{\partial(N\mathcal{H})}{\partial \sigma}=0 \;\;\rightarrow\;\; p_{\sigma}=const.\;,\label{h3}
\end{equation}
\begin{equation}
\dot{a}=\frac{\partial(N\mathcal{H})}{\partial p_a}=-\frac{Np_a}{12a}\;,\label{h4}
\end{equation}
\begin{equation}
\dot{T}=\frac{\partial(N\mathcal{H})}{\partial p_T}=\frac{N}{a}\;,\label{h5}
\end{equation}
\begin{equation}
\dot{\sigma}=\frac{\partial(N\mathcal{H})}{\partial p_{\sigma}}=\frac{N}{a^3}\quad. \label{h7}
\end{equation}
Let us choose $t=T$, so that $N=a$ in (\ref{h5}). Then, $T$ must be the conformal time, so that the role of time in the theory is realized by the radiation fluid. Using the gauge $N=a$, the Hamilton equations and the constraint equation (\ref{vinc}), we can obtain the equation of classical dynamic for the scale factor $a$
\begin{equation}
{\dot{a}}^2-\frac{1}{6}\big(p_T +\frac{p_{\sigma}}{a^{2}}\big)=0\;,\label{ceqa}
\end{equation}
or
\begin{equation}
\ddot{a}+\frac{1}{6}\frac{p_{\sigma}}{a^3}=0\;,\label{ceqa2}
\end{equation}

Note that comparing equation (\ref{ceqa}) with the Friedmann's equation obtained from Einstein's equation allow us to identify
\begin{eqnarray}
   p_\sigma = \frac{M_{\mathrm{Pl}}^{-2}}{2}\rho_\mathrm{s}^* \,, \\
	 p_T = \frac{M_{\mathrm{Pl}}^{-2}}{2}\rho_\mathrm{r}^* \,,
\end{eqnarray}
where $M_{\mathrm{Pl}}$ is the reduce Planck mass and $\rho_\mathrm{s}^*$ and $\rho_\mathrm{r}^*$ are the stiff matter and radiation densities when $a=a_*$.

We can now integrate equation (\ref{ceqa}) and find that
\begin{equation}
   \label{a_conforme}
   a(\tau)=\sqrt{rA^2t^2+2At} \,
\end{equation}
with $r=\frac{\rho_\mathrm{r}^*}{\rho_\mathrm{s}^*}$ and $A^2=\frac{M_{\mathrm{Pl}}^{-2}}{3}\rho_\mathrm{s}^*$.

\begin{figure}[ht]
\centering
\includegraphics[width=0.8\textwidth]{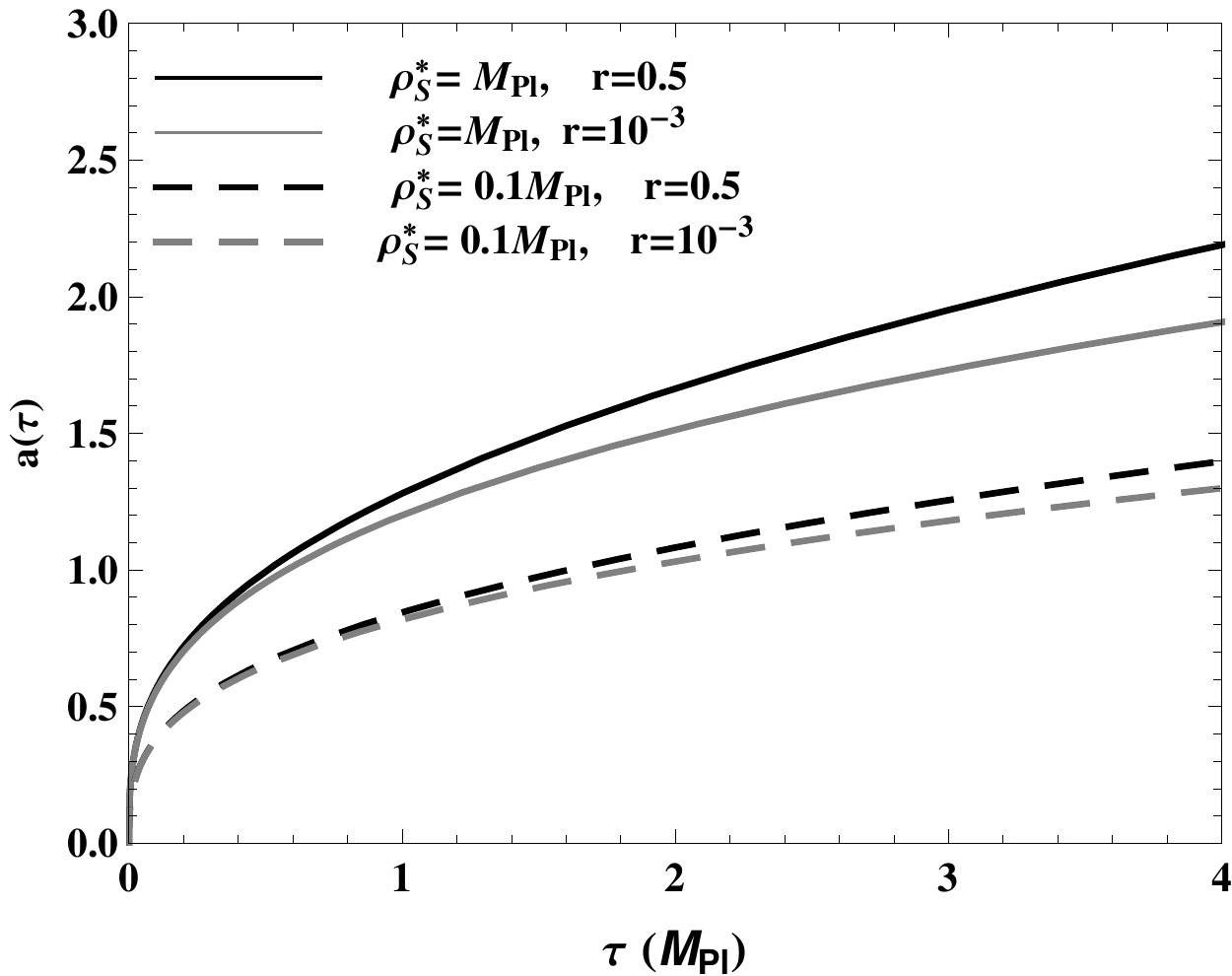}
\caption{Classical behavior of a two fluids model: stiff matter and radiation}
\label{classico}
\end{figure}

The conformal time $t$ is related to the cosmic time $\tau$ via $d\tau=N(t)dt=a(t)d\tau$ and with help of equation (\ref{a_conforme}) we find
\begin{equation}
   \label{t_cosmico}
	 \tau = \frac{\ln{\left(\sqrt{2rA^3}\right)}}{\sqrt{r^3A^2}}+\frac{a(\tau)}{2}\left(\frac{1}{rA}+t\right)-\frac{\ln{\left(rA^2t^{1/2} +\sqrt{2rA^3+r^2A^4t}\right)}}{\sqrt{r^3A^2}} \,.
\end{equation}
Using equations (\ref{a_conforme}) and (\ref{t_cosmico}) we can plot the scale factor $a$ as a function of the cosmic time $\tau$. In Figure \ref{classico} we show the classical behavior of the scale factor for different values of stiff matter and radiation densities at a fixed point $a_*=1$. The scale factor growth rate is more sensitive to the stiff matter quantity, but since it value falls very rapidly the scale factor will approach very fast the behavior of a universe filled with radiation. In this scenario,  the universe containing stiff matter and radiation is singular. 

Of course, for small values of $t$ e $a$, the classical formalism do not take into account the important quantum effects of this phase. It is therefore necessary to quantize the theory obtained.

\section{Quantum model}
\label{quant}

The quantum model is constructed making the quantification of canonically associated moments in equation (\ref{Hf001}) following the procedure proposed by Dirac \cite{Dirac}. The associated moments to $a$, $T$ e $\sigma$ coordinates are
\begin{displaymath}
\hat{p}_a=-i\frac{\partial}{\partial a}\;\;\;,\hat{p}_T=-i\frac{\partial}{\partial T}\;\;\;,\hat{p}_{\sigma}=-i\frac{\partial}{\partial \sigma}\;.
\end{displaymath}
Thus, the Hamiltonian constraint $\mathcal{H}$ becomes the $\hat{H}$ operator, which acts like the annihilator of the wave function of the universe $\Psi$. Then,
\begin{equation}
\left(-\frac{\partial^2}{\partial a^2}-144ka^2+\frac{24i}{a^{3\alpha-1}}\frac{\partial}{\partial t}+\frac{24i}{a^{3\beta-1}}\frac{\partial}{\partial \sigma}\right)\Psi(a,\sigma,t)=0\;.\label{2}
\end{equation}
The above equation have the form of a Schr\"odinger equation. This is a consequence of the Schutz's formalism used to describe the material content of the universe. Moreover, the scale factor of the universe is defined in the domain $(0,\infty]$. This means that the operator $\hat{H}$ is not, in general, self-adjoint. So in order to have a unitary evolution, the domain of $\hat{H}$ operator defined above must be: $D(\hat{H}):C_c^{\infty}(\mathbb{R}^{+}_*\times \mathbb{R}\times\mathbb{R}^+;\frac{\partial}{\partial a}\Psi_{(a=0)}=\alpha\Psi_{(a=0)})$. 

The function $\Psi$ can be write in the form of a stationary wave, i.e.,
\begin{equation}
\Psi(a,\sigma,t)=\psi(a,\sigma)e^{iEt}\;.\label{3}
\end{equation}
So
\begin{equation}
\left(-\frac{\partial^2}{\partial a^2}-144ka^2-\frac{24E}{a^{3\alpha-1}}+\frac{24i}{a^{3\beta-1}}\frac{\partial}{\partial \sigma}\right)\psi(a,\sigma)=0\;.\label{4}
\end{equation}

Separating variables, we can write
\begin{displaymath}
\psi(a,\sigma)=R(a)\Sigma(\sigma)\;.
\end{displaymath}
Then,
\begin{equation}
-\frac{1}{R(a)}\frac{\partial^2}{\partial a^2}R(a)-144ka^2-24Ea^{1-3\alpha}=-24ia^{1-3\beta}\frac{1}{\Sigma(\sigma)}\frac{\partial}{\partial \sigma}\Sigma(\sigma)\;.\label{5}
\end{equation}

We can adopt the solution of $\sigma$:
\begin{equation}
\Sigma(\sigma)=Ae^{-n\sigma}\;,\label{6}
\end{equation}
with $n\in\mathbb{C}$ and $A$ constant. Then
\begin{equation}
\frac{d^2R(a)}{da^2}+(144ka^2+24Ea^{1-3\alpha}+24ina^{1-3\beta})R(a)=0\;.\label{7}
\end{equation}

We restrict the analysis to $\alpha=1/3$, $\beta=1$ and $k=0$ case. With the solution of (\ref{7}), we can write the solution of (\ref{2}) for the wave function $\Psi$ as
\begin{equation}
\Psi_{En}(a,\sigma,t)=\sqrt{a}\left[CJ_{\frac{\sqrt{1-96in}}{2}}(\sqrt{24E}a)+DY_{\frac{\sqrt{1-96in}}{2}}(\sqrt{24E}a)\right]e^{-n\sigma+iEt}\;,\label{10}
\end{equation}
with $C$ and $D$ constants. Submitting the wave function to the boundary condition $\Psi(0,\sigma,t)=0$, we obtain
\begin{equation}
\Psi_{En}(a,\sigma,t)=C\sqrt{a}J_{\frac{\sqrt{1-96in}}{2}}(\sqrt{24E}a)e^{-n\sigma+iEt}\;.\label{11}
\end{equation}

Although boundary conditions, the above wave function is not square integrable. It is therefore necessary to construct the wave packets in $E$ and $n$. With this, the wave packet can be written in the form
\begin{equation}
\Psi_{En}(a,\sigma,t)=C\sqrt{a}\int dn\,g(n)e^{-n\sigma}\int dEf(E)J_{\frac{\sqrt{1-96in}}{2}}(\sqrt{24E}a)e^{iEt}\;.\label{12}
\end{equation}

If we do $\sqrt{24E}=q$ and make the choice \begin{displaymath}
f(q)=12q^{\frac{\sqrt{1-96in}}{2}}e^{-\gamma q^2}\;,\,\,\,\gamma\in\mathbb{R}\quad,
\end{displaymath}
we can construct the wave packet in $E$, which is written in terms of exponential functions
\begin{equation}
\Psi_{qn}(a,\sigma,t)=\frac{C\,\sqrt{a}}{12}\int dn\,g(n)e^{-n\sigma}\int dq\,q^{\frac{\sqrt{1-96in}}{2}+1}J_{\frac{\sqrt{1-96in}}{2}}(qa)e^{(-\gamma+\frac{it}{24})q^2}\;.\label{13}
\end{equation}
The above integral can be solved analytically e gives \cite{abra}
\begin{equation}
\Psi_n(a,\sigma,t)=\frac{C\sqrt{a}}{2\gamma-\frac{it}{12}}e^{-\frac{a^2}{4\gamma-\frac{it}{6}}}\int dn g(n)e^{-n\sigma}\left(\frac{a}{2\gamma-\frac{it}{12}}\right)^{\frac{\sqrt{1-96in}}{2}}\;,\label{14}
\end{equation}
subject to the following conditions $Re(\gamma-\frac{it}{24})>0$ e $Re(\frac{\sqrt{1-96in}}{2})>-1$. The above integration can also be solved analytically if we do $\frac{1-96in}{4}=u^2$ and $g(n)\rightarrow h(u)=-\frac{e^{-\xi u^2}}{2i\sqrt{6\pi}Cu}$, $\xi\in\mathbb{R}$. So, we obtain
\begin{eqnarray}
\Psi(a,\sigma,t)&=&\frac{6\sqrt{a}}{(24-it)\sqrt{24\xi-i\sigma}}\left(\frac{12ia}{t+24i\gamma}\right)^{\frac{12}{24\xi-i\sigma}}e^{\frac{i\sigma}{96}-\frac{6a^2}{24\gamma-it}}\Bigg{[}1 + \nonumber \\
&+& \mathrm{erf}\left(\sqrt{\frac{6}{24\xi-i\sigma}}\mathrm{ln}\left(\frac{12ia}{t+24i\gamma}\right)\right)\Bigg{]}   \;,\label{16}
\end{eqnarray}
where $\mathrm{erf}(x)$ is the error function (see Appendix). This type of function has no closed form and their values for different arguments must be obtained numerically.

\begin{figure}[ht]
\centering
\includegraphics[width=0.45\textwidth]{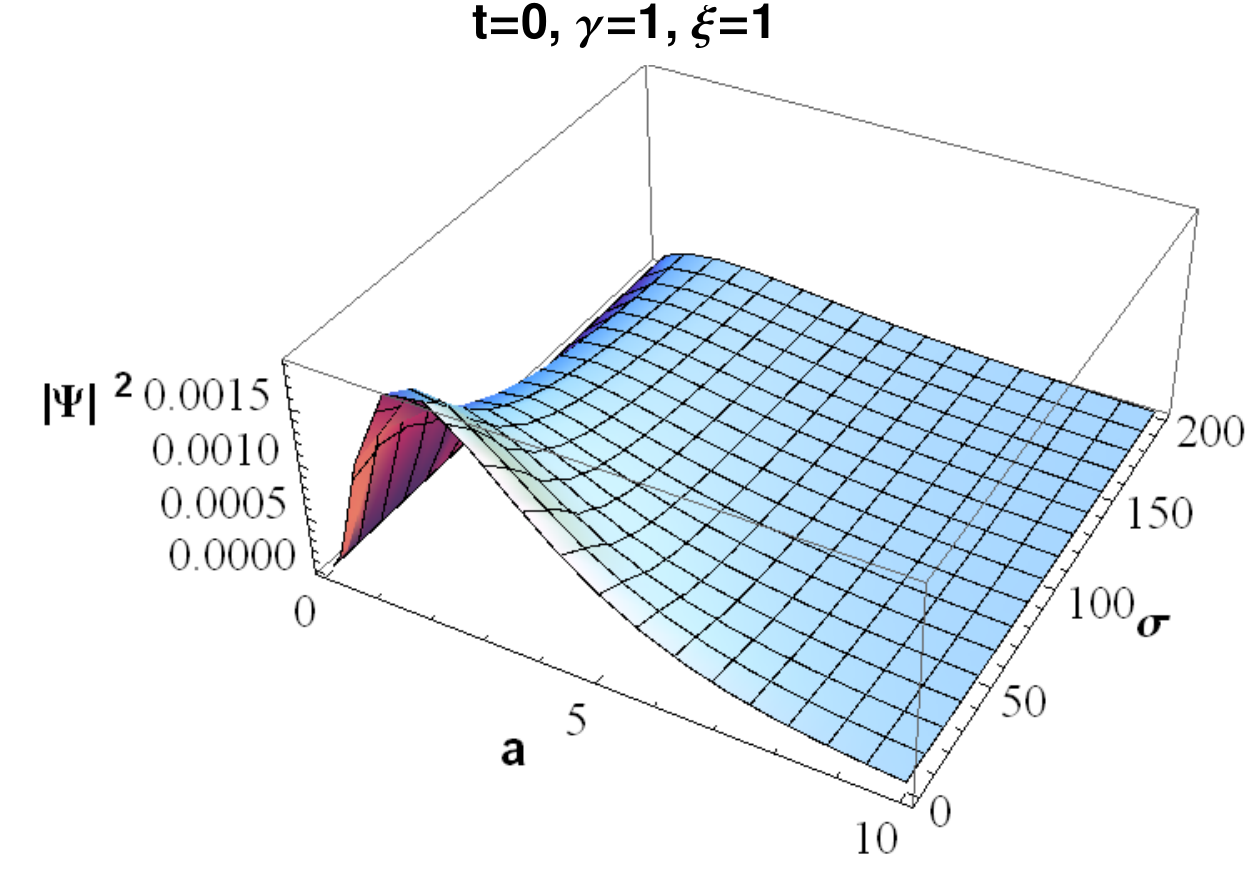}
\includegraphics[width=0.45\textwidth]{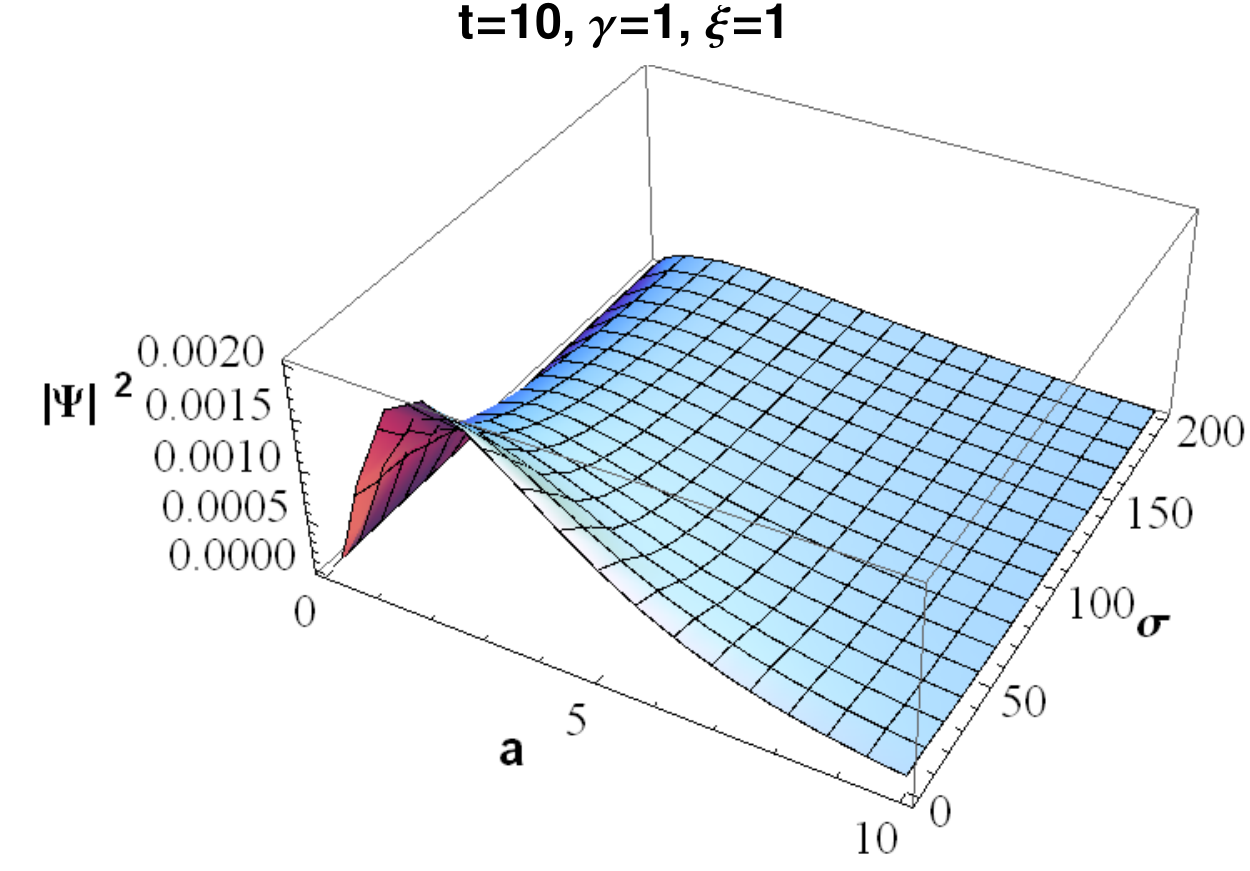}
\\
\includegraphics[width=0.45\textwidth]{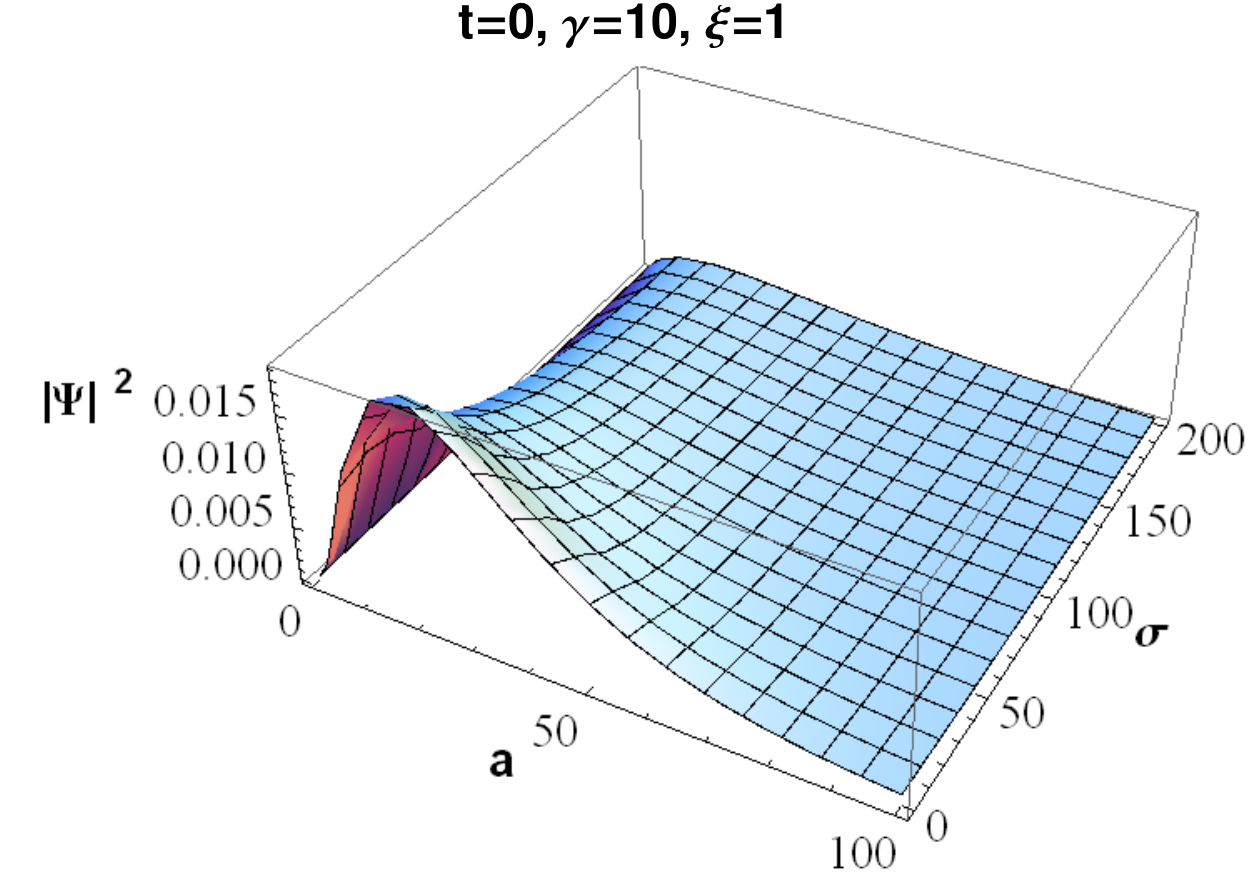}
\includegraphics[width=0.45\textwidth]{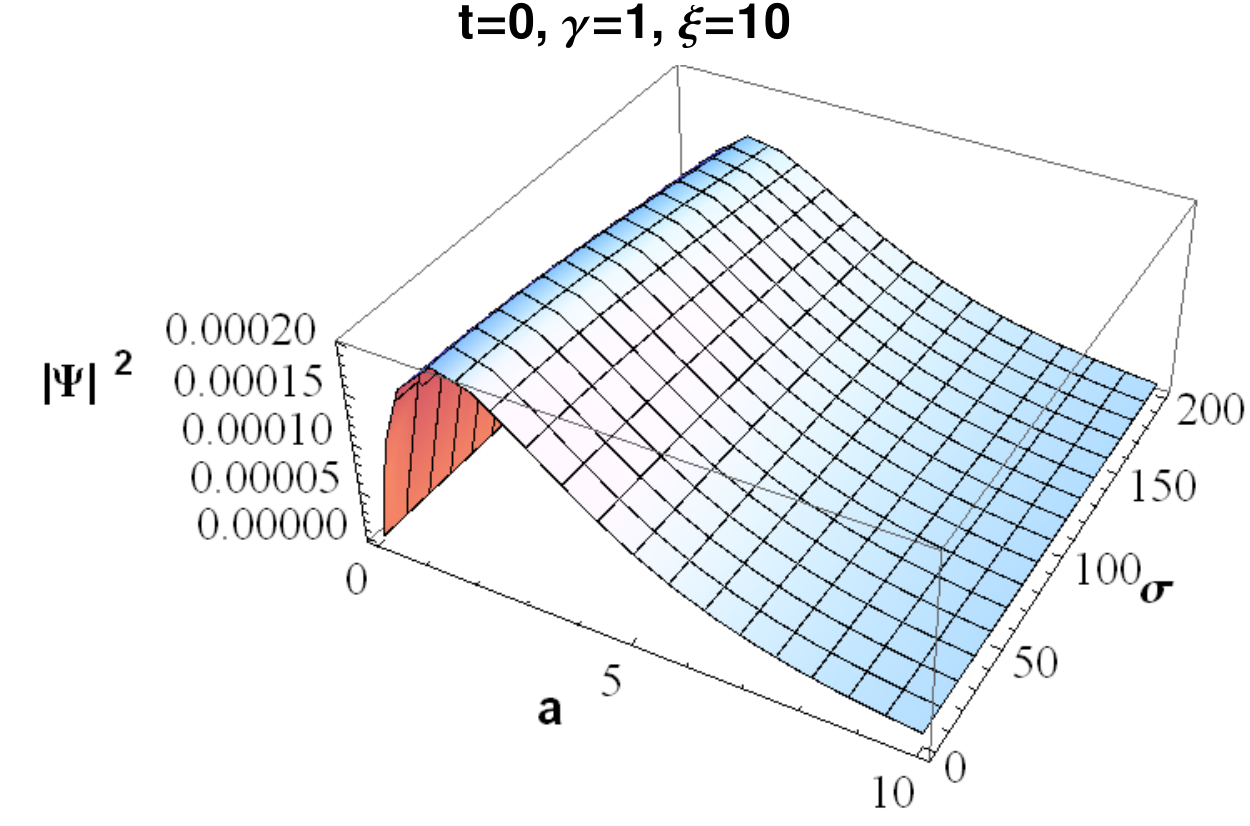}
\caption{The unormalized wave packet in the $a \times \sigma$ plane for fixed $t$ with various combinations of the free parameters $\gamma$ and $\xi$.}
\label{psi_a_sigma}
\end{figure}

\begin{figure}[ht]
\centering
\includegraphics[width=0.45\textwidth]{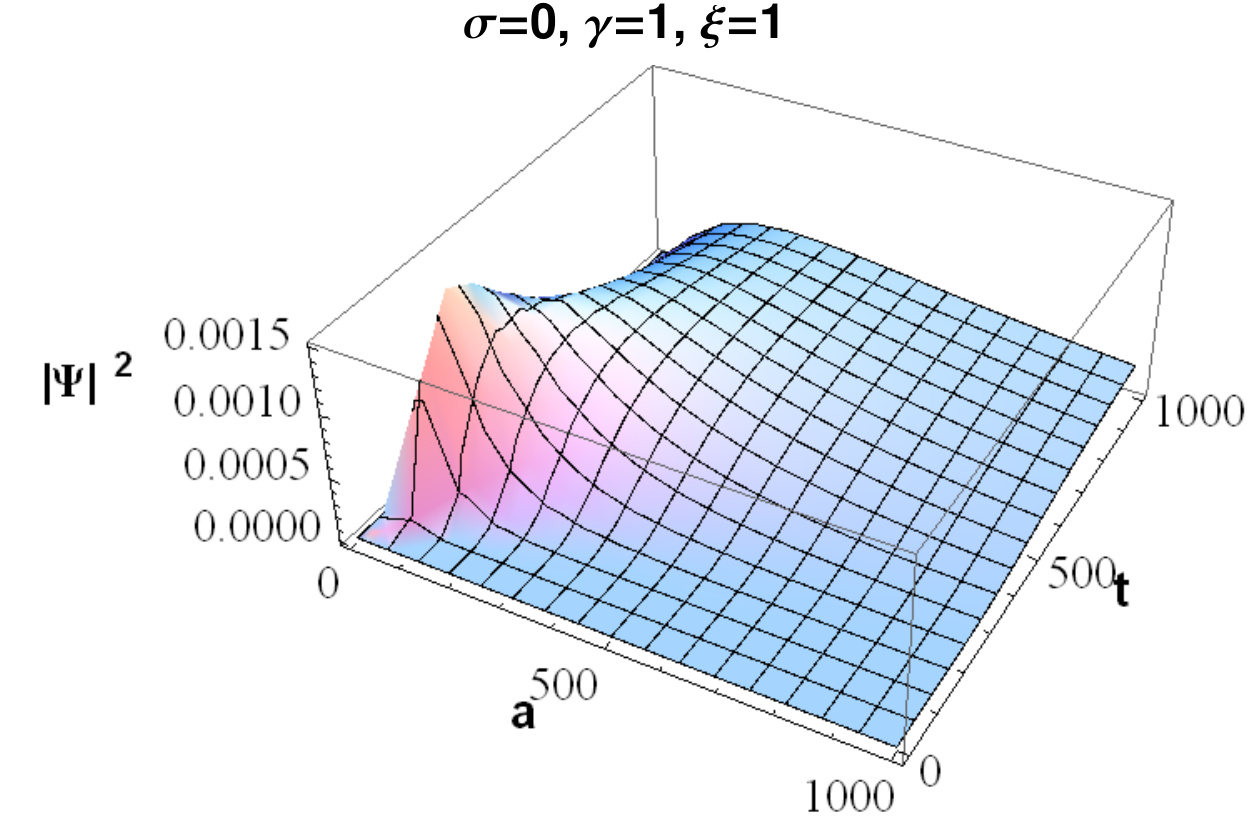}
\includegraphics[width=0.45\textwidth]{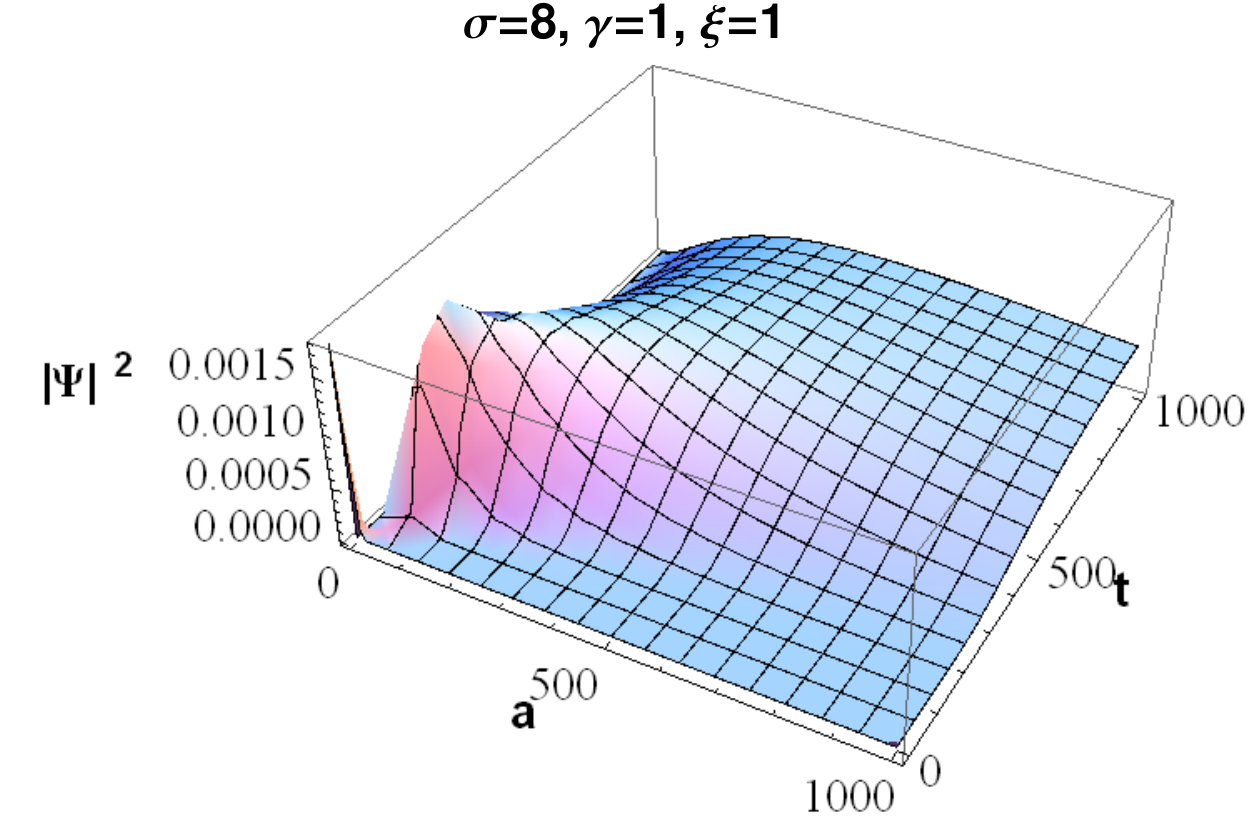}
\\
\includegraphics[width=0.45\textwidth]{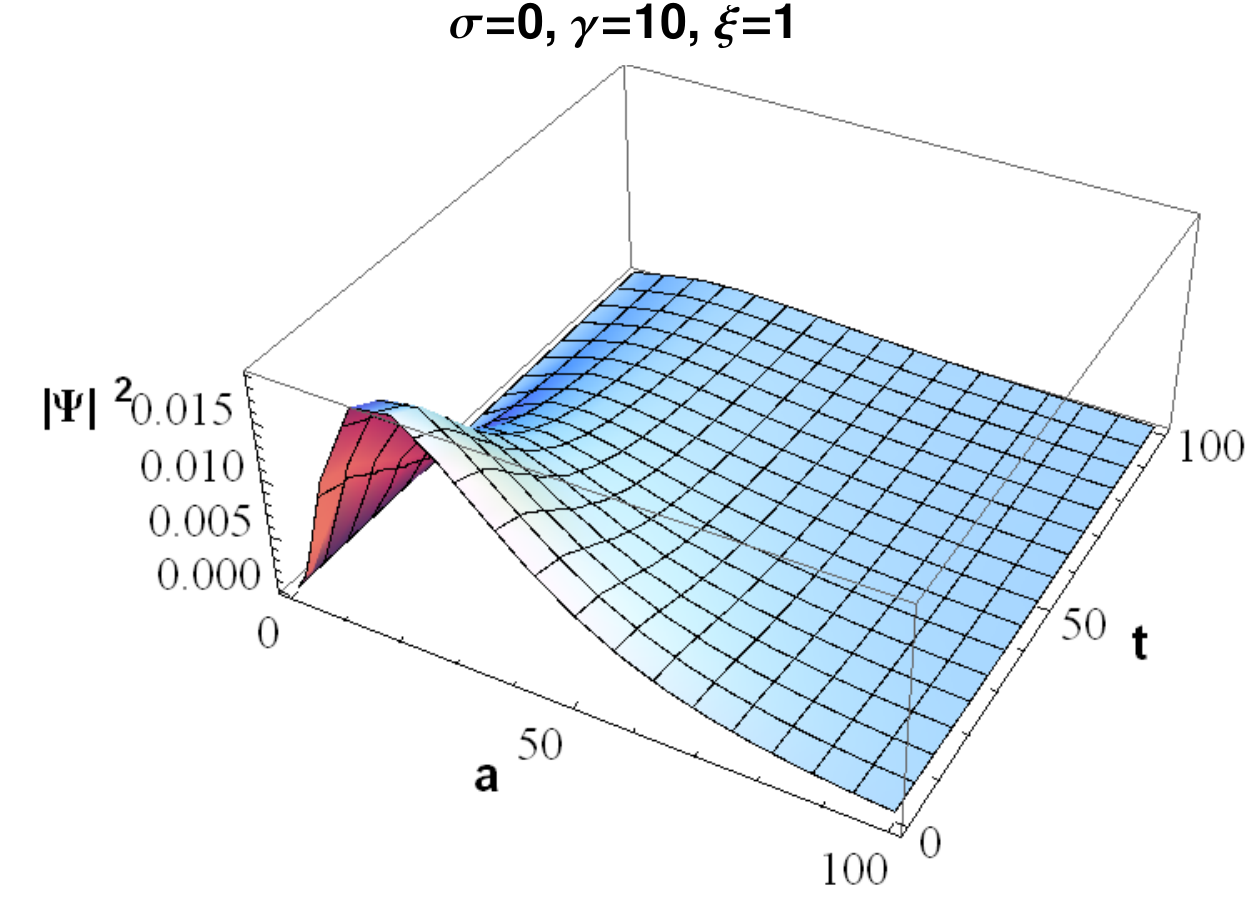}
\includegraphics[width=0.45\textwidth]{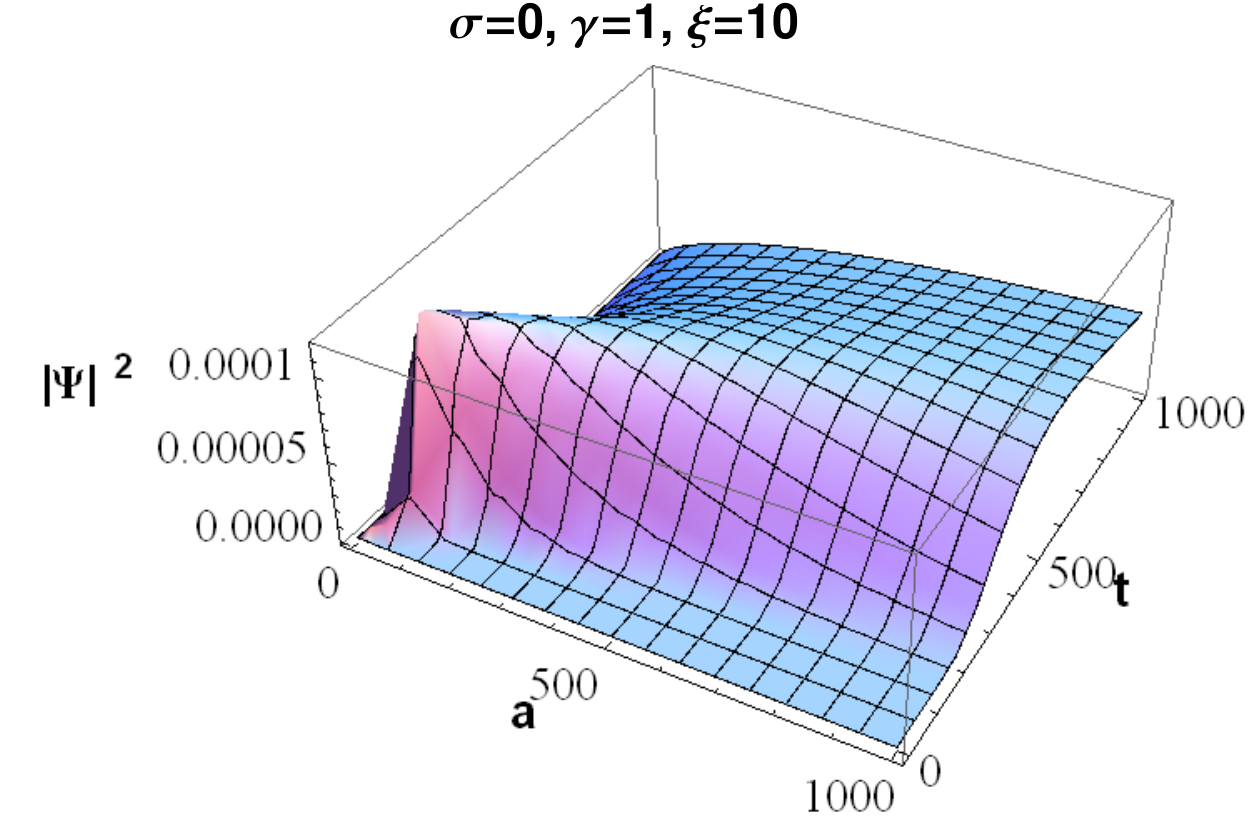}
\caption{The unormalized wave packet in the $a \times t$ plane for fixed $\sigma$ with various combinations of the free parameters $\gamma$ and $\xi$.}
\label{psi_a_t}
\end{figure}

According to the Figures \ref{psi_a_sigma} and \ref{psi_a_t}, it is easy to verify that $\Psi(a,\sigma,t)\rightarrow 0$ when the scale factor vanish, $a\rightarrow 0$, which is consistent with the imposed boundary conditions. In this case, defining the probability density  $P(a)=|\Psi(a)|^2$, it is noted that $P(a)\rightarrow0$ when $a\rightarrow0$ \cite{Papa} so that the singularity, observed in the classical version, is avoided at this quantum model.

\begin{figure}[ht]
\centering
\includegraphics[width=0.45\textwidth]{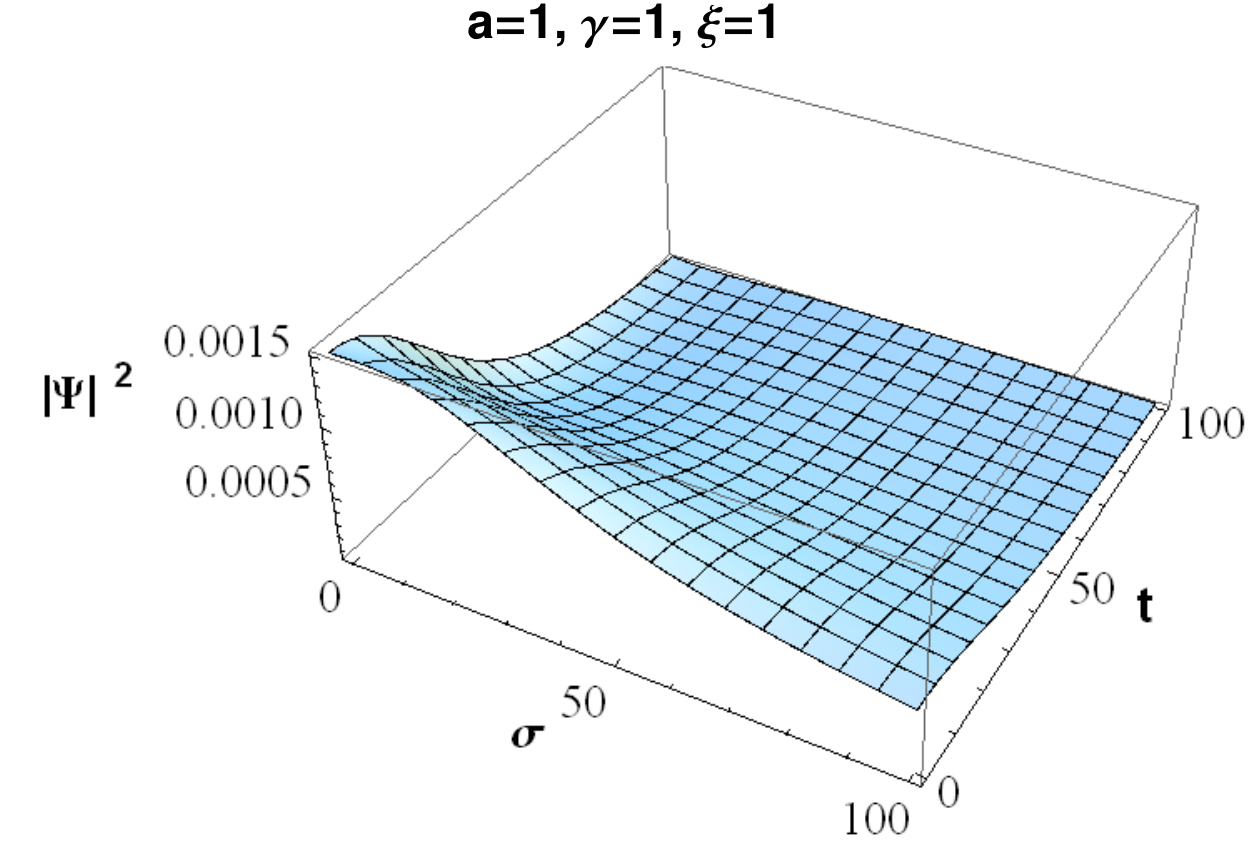}
\includegraphics[width=0.45\textwidth]{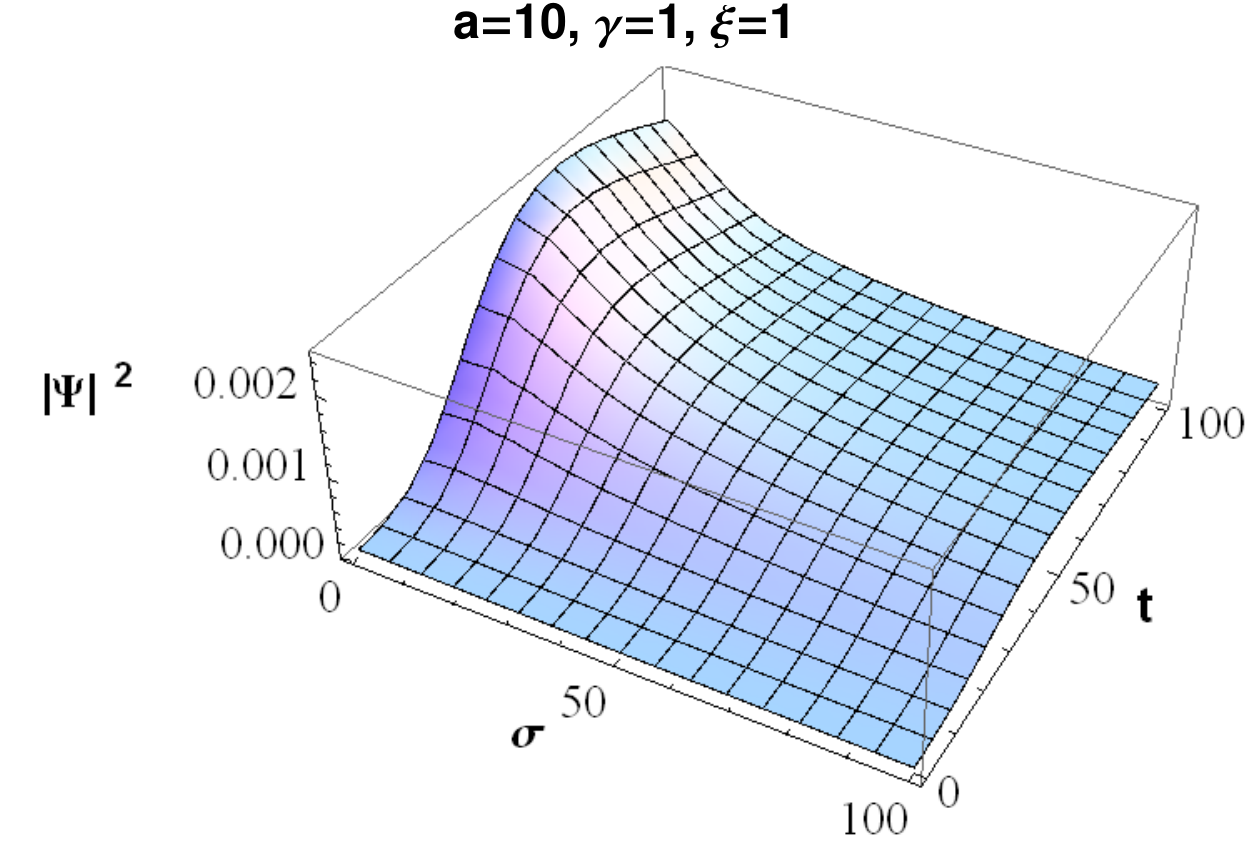}
\\
\includegraphics[width=0.45\textwidth]{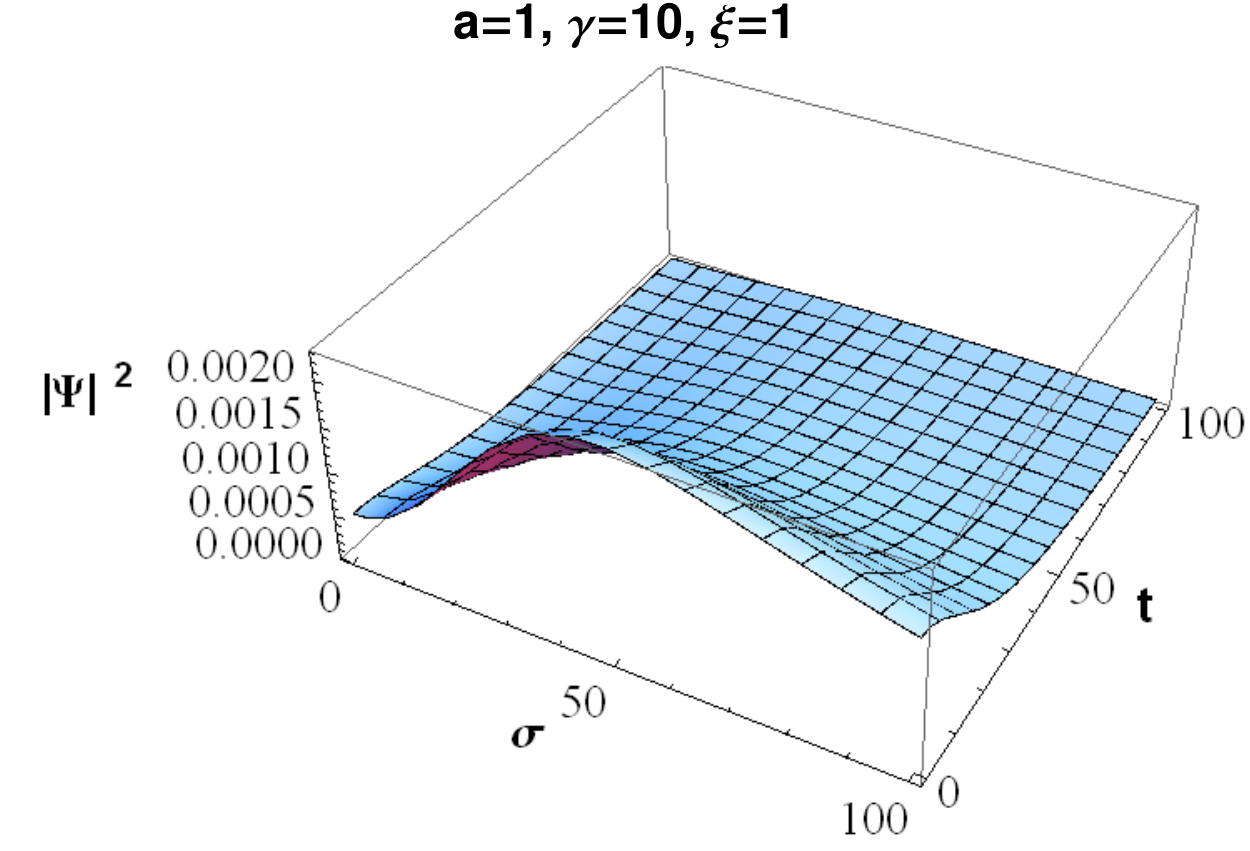}
\includegraphics[width=0.45\textwidth]{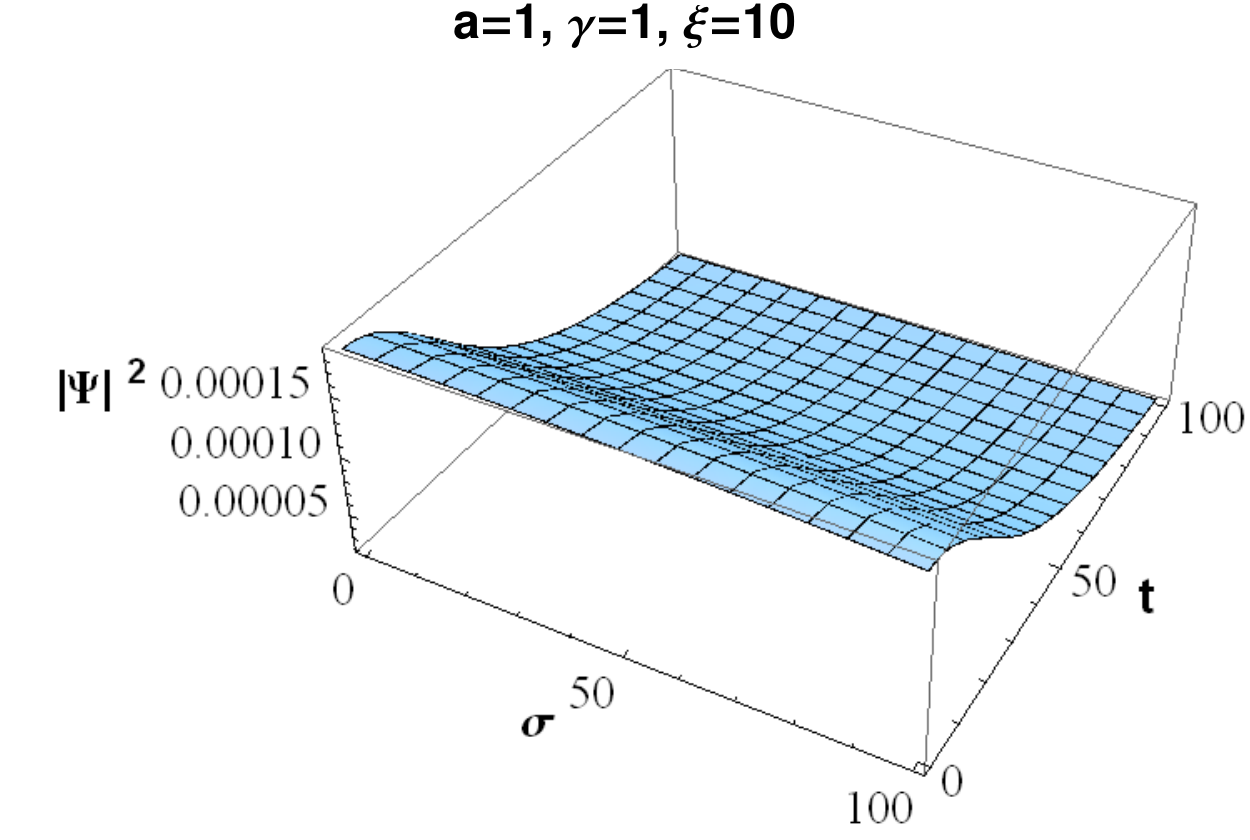}
\caption{The unormalized wave packet in the $\sigma \times t$ plane for fixed $a$ and with various combinations of the free parameters $\gamma$ and $\xi$.}
\label{psi_sigma_t}
\end{figure}

\begin{figure}[ht]
\centering
\includegraphics[width=0.45\textwidth]{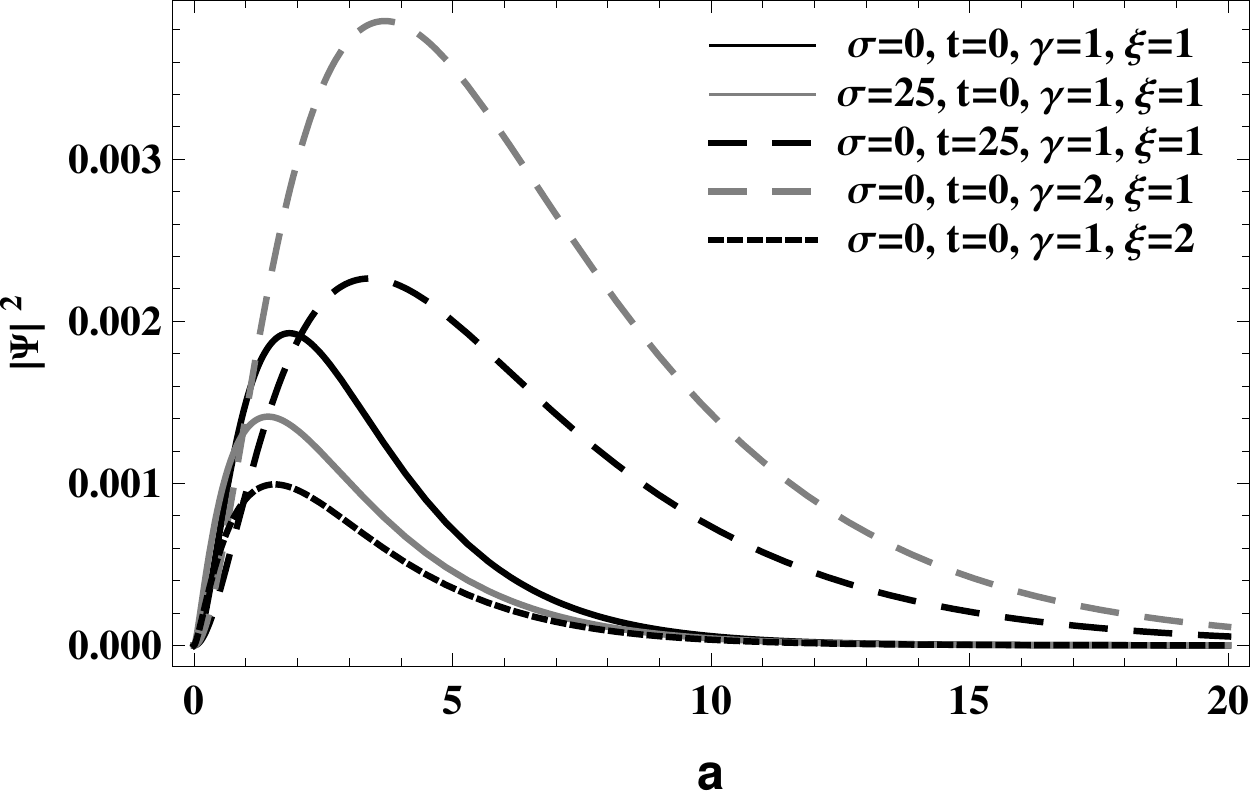}
\includegraphics[width=0.45\textwidth]{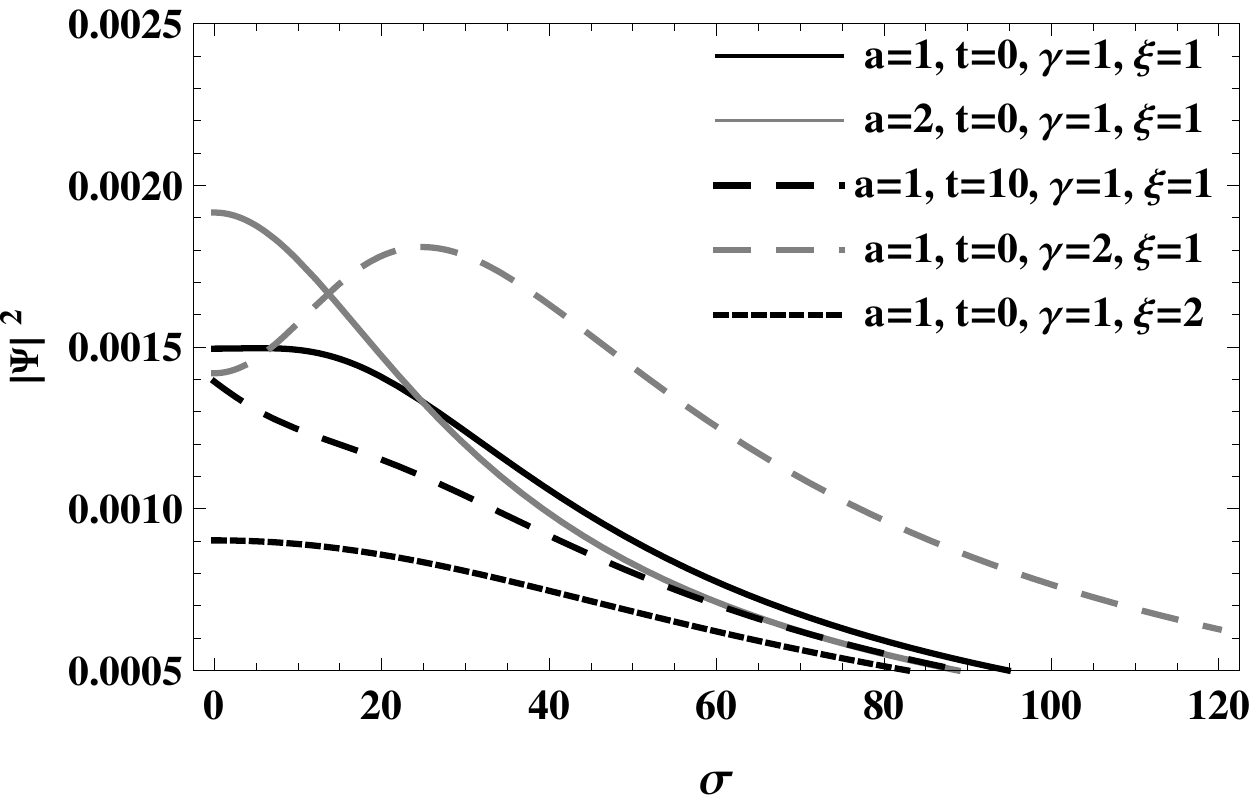} \\
\includegraphics[width=0.45\textwidth]{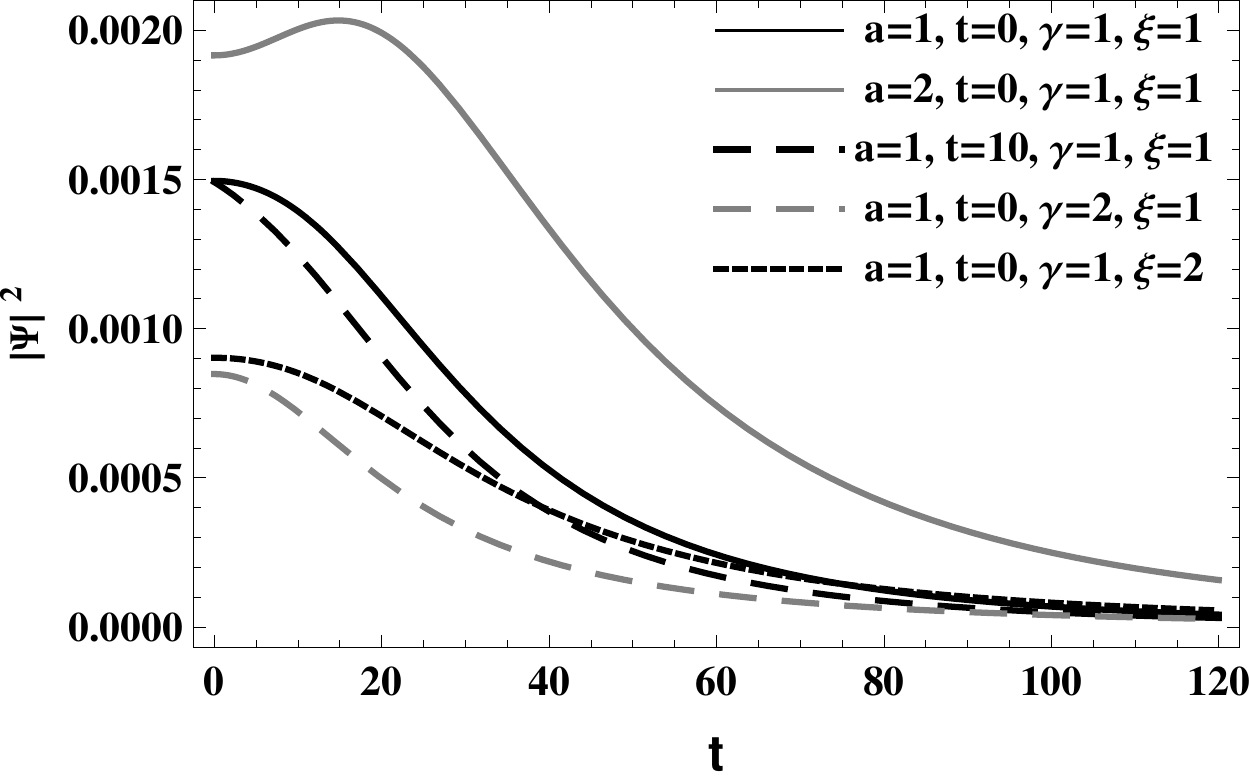}
\caption{The unormalized wave packet as a function of only one of the three variables, with the other two fixed and for various combinations of the free parameters $\gamma$ and $\xi$.}
\label{psi_1dim}
\end{figure}

\begin{figure}[ht]
\centering
\includegraphics[width=0.45\textwidth]{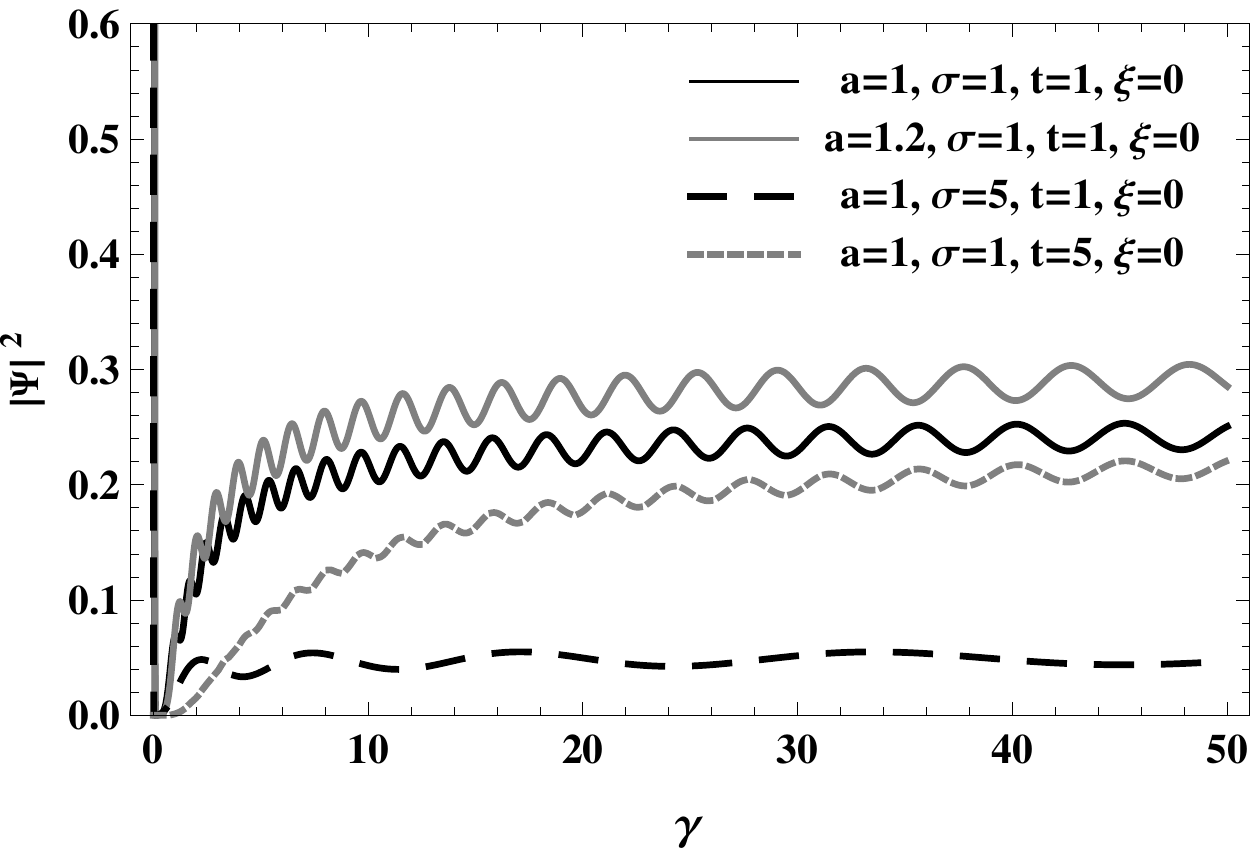}
\includegraphics[width=0.45\textwidth]{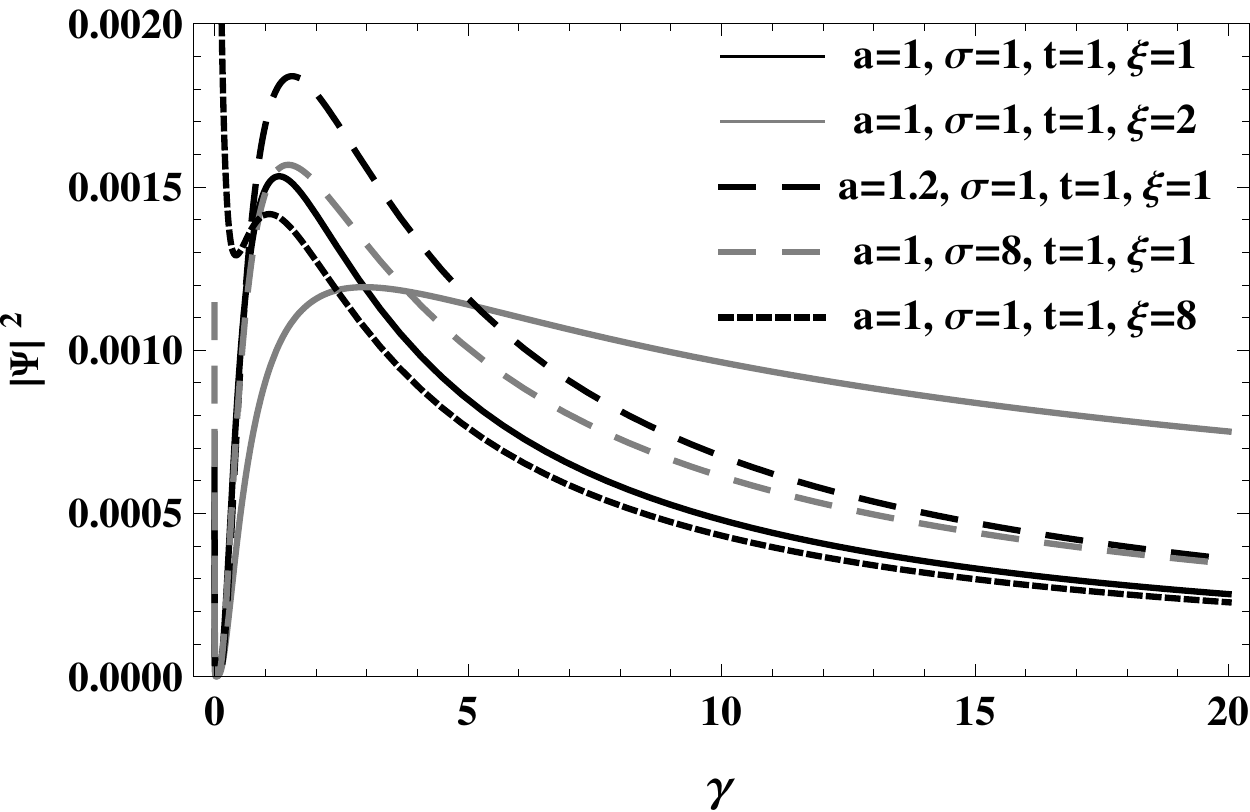} \\
\includegraphics[width=0.45\textwidth]{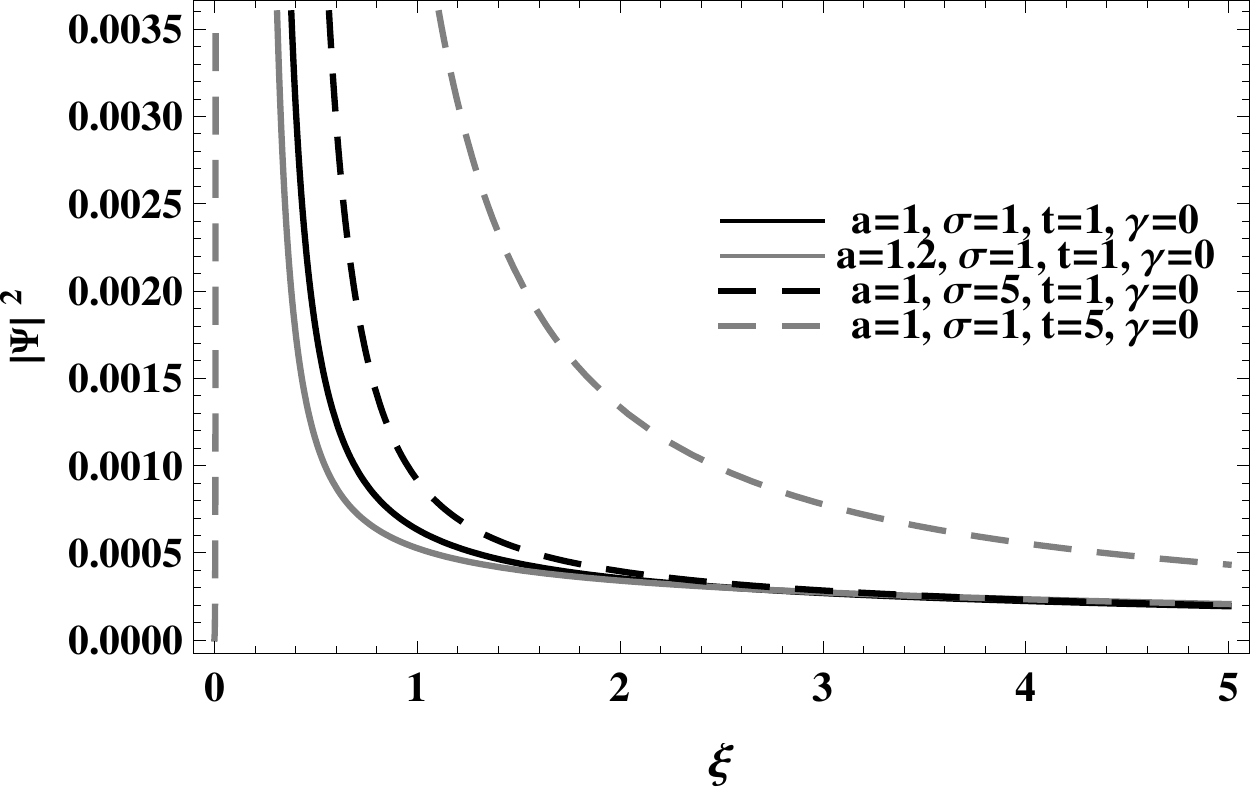}
\includegraphics[width=0.45\textwidth]{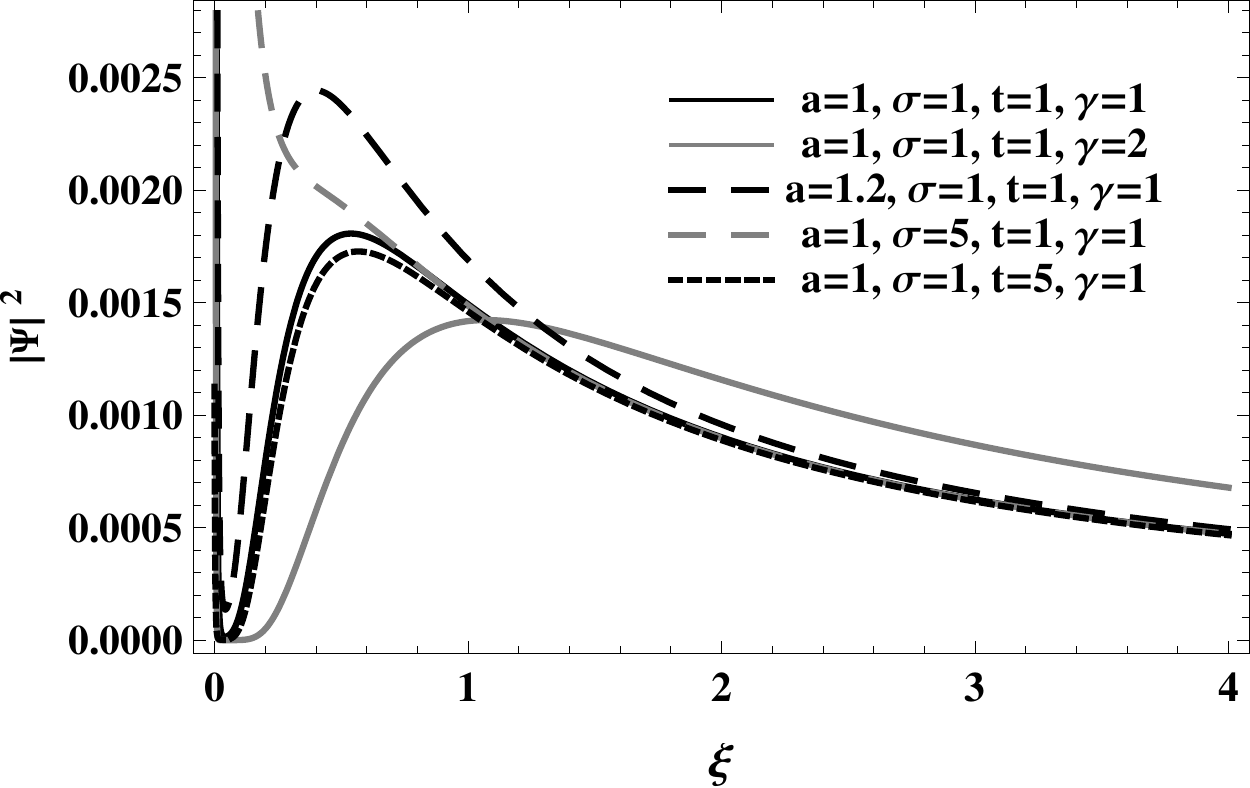}
\caption{The dependence of the unormalized wave packet in relation to the free parameters $\gamma$ and $\xi$, for fixed values of the variables $a$, $\sigma$ and $t$.}
\label{psi_1dim_gamma_xi}
\end{figure}

\section{Conclusions}
\label{concl}
In this paper, we study the quantum FRW model with stiff matter and radiation and we solve the Wheeler-DeWitt equation in minisuperspace to obtain the wave function of the corresponding universe. The perfect fluid is described by the Schutz's canonical formalism, which allows to attribute dynamical degrees of freedom to matter. 

In the classical model, the universe of null curvature  expands forever from an initial singularity, as can be seen in equations (\ref{a_conforme}), (\ref{t_cosmico}) and in Figure \ref{classico}. Indeed,  this singular point can be avoided  in quantum cosmology. In quantum treatment the model of fluids was investigated on the basis of the associated Wheeler-DeWitt equation. The introduction of a time variable phenomenologically using the  degrees of freedom of radiation fluid allows to obtain a Hamiltonian constraint linear in one of the momenta and the Wheeler-DeWitt equation can be reduced to a Schr\"odinger like equation, whose solution is the wave function of the universe. Applying the boundary conditions, this function was explicitly written and the wave packets was constructed analytically.

The behavior of the wave packet (\ref{16}) is showed in Figures \ref{psi_a_sigma}, \ref{psi_a_t} and \ref{psi_sigma_t} for each two pairs of variables $a$, $\sigma$ and $t$. The packet goes to $0$ as the variables grow, but the decaying rate depends on the values of the third fixed variable and the free parameters $\gamma$ and $\xi$. It is worth to note that in equation (\ref{16}) the variable $t$ appears many times accompanied by the parameter $\gamma$ and $\sigma$ by $\xi$. This fact can be used as a manner to regulate how fast the packet approaches zero for the directions $\sigma$ and $t$. In Figure \ref{psi_1dim} we show the cross section of the wave packet and see how different combinations of values for the variables and the parameters can change the intensity and position of the peak, as the decaying rate. The dependence in relation to the free parameter is showed in Figure \ref{psi_1dim_gamma_xi}.

The behavior of the wave packet, shown in Figures (\ref{psi_a_sigma}) and (\ref{psi_a_t}), makes it clear that $\Psi(a,\sigma,t)$ tends to zero when $a$ tends to zero. Because of this, the probability density $P(a)$ tends to zero when $a$ tends to zero, so that the classical singularity is avoided, unlike what happens in the classical case. 

As perspective to future work, the investigation of new scenarios containing other types of fluid is the natural way. It is hoped that this will highlight the role played by new degrees of freedom added to cosmological models, as well as contribute to the discussion about the quantization procedures applied to cosmology.


\section*{Acknowledgement}

\noindent
This work has received  financial supporting from FAPES (Brazil). 


\begin{appendices}

\section{The error function}
\label{apendice}

Error function $\mbox{erf}~x$ is given as \cite{abra}
\begin{equation}
\mbox{erf}~x = \frac{2}{\sqrt{\pi}}\int^x_0~e^{-t^2}~dt\quad.
\end{equation}

There is a direct connection between the error function and the Gaussian function and the normalized Gaussian function, that are given by
\begin{eqnarray}
f(x) &=& a~e^{\frac{-(x - b)^2}{2\sigma^2}}\quad,\nonumber \\
g(x) &=& \frac{1}{\sqrt{2\pi}}~e^{\frac{-(x - b)^2}{2\sigma^2}}\quad,
\end{eqnarray}
respectively, where $a, b$ are constants and $\sigma$ is the standard deviation, sometimes called the Gaussian RMS width. The graph of a Gaussian is a characteristic symmetric bell curve shape with the parameter $a$ being the height of the curve's peak, $b$ is the position of the center of the peak and $\sigma$ controls the width of the bell.
\par
With the substitution
\begin{eqnarray}
t^2 &=& \frac{(x - b)^2}{2\sigma^2}\quad,\nonumber \\
dt &=& \frac{1}{\sqrt 2 \sigma}~dx \quad,
\end{eqnarray}
we can integrated the normalized Gaussian function between $-x$ and $+x$ and obtain
\begin{equation}
\int^{+x}_{-x}~g(x)~dx = \frac{2}{\sqrt\pi}\int^x_ 0~e^{-t^2}~dt = \mbox{erf}~x\quad,
\end{equation}
since the normalized Gaussian is symmetric about the $y$-axis.
\par
Some properties and approximations of the error function are
\begin{eqnarray}
\mbox{erf}~(-x) &=& -\mbox{erf}~x\quad,\\
\mbox{erf}~0 &=& 0\quad,\\
\mbox{erf}~(x^*) &=& (\mbox{erf}~x)^*~~~~\mbox{where}~^*~\mbox{denotes complex conjugation}\quad,\\
\mbox{erf}~\infty &=& 1\quad,\\
\mbox{erf}~(-\infty) &=& -1\quad,\\
\mbox{erf}~x &=& \frac{2}{\sqrt\pi}\int^x_0~\sum_{n = 0}^{\infty}~\frac{(-1)^n t^{2n}}{n!}~dt~~~~\mbox{for}~x << 1\quad,\\
\mbox{erf}~x &=& \frac{2}{\sqrt\pi}~\sum_{n = 0}^{\infty}~\frac{(-1)^n x^{2n + 1}}{(2n + 1)n!}\quad,\nonumber\\
             &=& \frac{2}{\sqrt\pi}\biggl(x - \frac{x^3}{3} + \frac{x^5}{10} - \frac{x^7}{42} + ...\biggr)~~~~\mbox{for}~x << 1\quad,\\
\mbox{erf}~x &=& 1 - \frac{e^{-x^2}}{\sqrt\pi~x}\biggl(1 - \frac{1}{2x^2} - ...\biggr)~~~~\mbox{for}~x >> 1\quad.
\end{eqnarray}
\par
The complementary error function, commonly denoted erfc~$x$, is defined by
\begin{eqnarray}
\mbox{erfc}~x &=& 1 - \mbox{erf}~x\quad,\nonumber\\
              &=& \frac{2}{\sqrt\pi}\int_x^{\infty}~e^{-t^2}~dt\quad.
\end{eqnarray}
\par
Some special values, approximations and results of erfc~$x$ are
\begin{eqnarray}
\mbox{erfc}~(-x) &=& 2 -\mbox{erfc}~x\quad,\\
\mbox{erfc}~\infty &=& 0\quad,\\
\mbox{erfc}~(-\infty) &=& 2\quad,\\
\mbox{erfc}~0 &=& 1\quad,\\
\mbox{erfc}~x &=& \frac{\Gamma(1/2,x^2)}{\sqrt\pi}\quad,~~~~\mbox{where}~\Gamma(a,y)~\mbox{is the gamma function}\\
\int_0^{\infty}~\mbox{erfc}~x~dx	&=&	\frac{1}{\sqrt\pi}\quad.
\end{eqnarray}
\par
\begin{figure}[ht]
\centering
\includegraphics[width=0.8\textwidth]{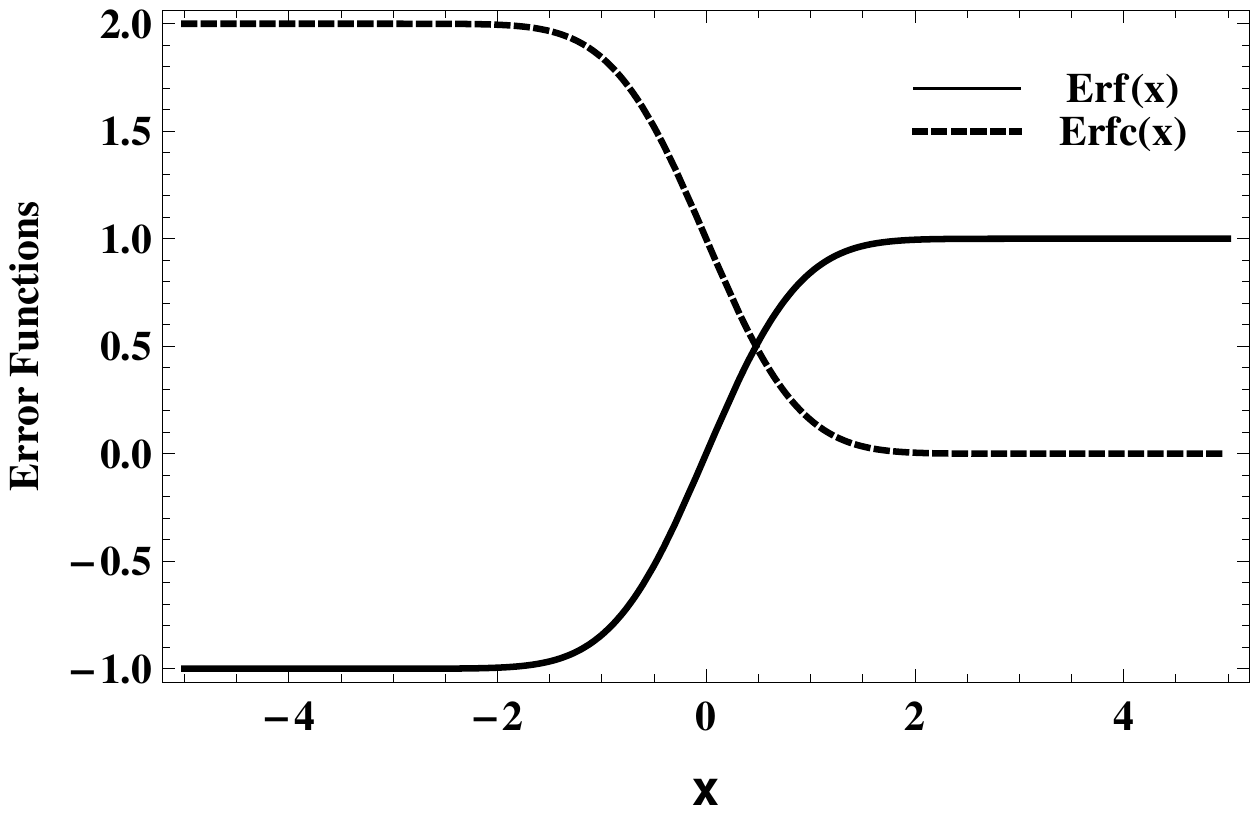}
\caption{The error and the complementary error functions.}
\label{fig:figure}
\end{figure}

We show in Figure (\ref{fig:figure}) the behaviour of the special functions treated here in this appendix.

\end{appendices}

\end{document}